\documentclass[conference]{IEEEtran}
\IEEEoverridecommandlockouts
\usepackage{cite}
\usepackage{amsmath,amssymb,amsfonts}
\usepackage{algorithmic}
\usepackage{graphicx}
\usepackage{textcomp}
\usepackage{xcolor}
\usepackage{subcaption}
\def\BibTeX{{\rm B\kern-.05em{\sc i\kern-.025em b}\kern-.08em
T\kern-.1667em\lower.7ex\hbox{E}\kern-.125emX}}
\begin{document}

    \title{Safety-Critical Lane-Change Control for CAV Platoons in Mixed Autonomy Traffic Using Control Barrier Functions\\
    \thanks{Identify applicable funding agency here. If none, delete this.}
    }

    \author{\IEEEauthorblockN{Fengqing Hu}
    \IEEEauthorblockA{\textit{Division of EMIA} \\
    \textit{HKUST}\\
    Hong Kong SAR, China \\
    fengqing.hu@connect.ust.hk}
    \and
    \IEEEauthorblockN{Huan Yu}
    \IEEEauthorblockA{\textit{INTR Thrust, Systems Hub} \\
    \textit{HKUST (Guangzhou)}\\
    Guangzhou, China \\
    Department of Civil and Environmental Engineering\\
    \textit{HKUST}\\
    Hong Kong SAR, China \\
    huanyu@ust.hk}
    }

    \maketitle

    \begin{abstract}
        Platooning can serve as an effective management measure for connected and autonomous vehicles (CAVs) to ensure overall traffic efficiency. Current study focus on the longitudinal control of CAV platoons, however it still remains a
        challenging problem to stay safe under lane-change scenarios where both longitudinal and lateral
        control is required. In this paper, a safety-critical control method is proposed conduct
        lane-changing maneuvers for platooning CAVs using Control Barrier Functions (CBFs). The proposed method is
        composed of two layers: a higher-level controller for general lane change decision control and a lower-level
        controller for safe kinematics control. Different from traditional kinematics controllers, this lower-level controller conducts not only
        longitudinal safety-critical control but also critically ensures safety for lateral control during the platooning
        lane change. To effectively design this lower-level controller, an optimization problem is solved with constraints
        defined by both CBFs and Control Lyapunov Functions (CLFs). A traffic simulator is used to conduct numerical traffic simulations in four safety-critical scenarios and showed the effectiveness of the proposed controller.
    \end{abstract}

    \begin{IEEEkeywords}
        Control barrier functions, platoon, lane change, connected and autonomous vehicles, mixed autonomy traffic
    \end{IEEEkeywords}

    \section{Introduction}
    With the rapid development of automated driving systems (ADS) and communication technologies, the control of connected and autonomous vehicles (CAVs) have been widely discussed in the recent years. Compared to human-driven vehicles (HDVs) and single autonomous vehicle without vehicle-to-vehicle (V2V) communication, CAVs can improve the efficiency of the transportation system by designing cooperative control strategies \cite{Sharan:2021}. In the literature, platooning has been applied in many studies to improve traffic efficiency by reducing the headways among the vehicles in the platoon \cite{Nieuwenhuijze:2012}\cite{Bian:2018}\cite{Bian:2019}. For platoon control, most existing studies focus on the longitudinal control and operations \cite{Li:2017} \cite{WangC:2020}. Adaptive cruise control (ACC) have been studied since an early time \cite{Marsden:2001}. It provides an automatic longitudinal control to keep a reasonable spacing gap from the preceding vehicle. Facilitated by V2V communication technologies, a smoother car following control strategy was achieved by connected and autonomous cruise control (CACC) where only longitudinal interactions between CAVs are considered \cite{Nieuwenhuijze:2012} \cite{Wang:2020}. Despite the boom of ADS and advanced driver assistance system (ADAS), it is anticipated that mixed autonomy traffic will still exist for a long period in future transportation systems. Some studies came to consider the longitudinal interaction between CAVs and HDVs. The leading cruise control (LCC), for instance, provided a longitudinal control strategy for a CAV to follow the preceding HDVs while leading the HDVs behind it \cite{Wang:2021}.
    
    Although longitudinal control plays an important role in platoon operations, scenarios requiring lateral control are also very common in the real transportation environment, for example, the lane change scenarios. Platoon operations considering lane change have been well investigated in a number of typical scenarios. In \cite{Rajamani:2000}, a strategy for CAV platoon operation in merge and diverge scenario was proposed. Two maneuvers are applied in the proposed strategy: split and join. When a CAV is changing its lane, the split maneuver will be conducted by the following CAV to leave sufficient spacing gap for the lane change CAV. After the lane change, the platoon comes to the join maneuver, by which the headway is reduced to improve traffic efficiency. \cite{Schmidt:2017} extended CACC to a  typical lane change scenario. A controller was designed to follow the predecessor vehicles in both the current lane and the target lane to keep a safe distance for lane change. \cite{Zhang:2022} designed an cooperative lane change control strategy for a CAV platoon based on machine learning and model predictive control (MPC), where several typical merge operations are considered. In this paper, we focus on the lane-change control of a CAV platoon in safety-critical scenarios.
    
    Safety is of crucial significance in CAV control. In mixed autonomy traffic, CAVs have to interact with HDVs and may face safety problems when HDVs take irregular or even extreme behaviors. Currently, a safe controller is usually achieved by solving optimization problems with safety-related constraints \cite{Zhu:2019} \cite{Zhai:2019}. In this paper, we will employ control barrier function (CBF) to design the safety constraint for CAV control. Restricted by CBFs, the dynamics system can guarantee not only safety, but also a progressive approaching speed to the barrier \cite{Ames:2019}. Combined with CBFs and control Lyapunov functions (CLFs), an optimization-based controller can realize multi-objective nonlinear control where both stability and safety constraints are both required \cite{Ames:2013}. CBF-based method has now been widely used in a number of control systems: robot control \cite{Nguyen:2016}, single automated vehicle control \cite{Ames:2014} \cite{Xiao:2022} \cite{He:2021} and longitudinal platoon control \cite{He:2018}. However, to the best of our knowledge, safety-critical control with CBFs in lateral platoon operations has not been well studied.
    
    The remainder of this paper is organized as follows. Section II formulates the overall safety-critical control problem and provides the details of the designed controller. Section III presents a series of comparative experiments to verify the proposed method.

    \section{Problem Formulation}

    To successfully conduct lane change for CAV platoons, two platoon operations are designed. The first step for lane change is to split the platoon. One of the CAVs aims to change lane and leave the platoon, while the following CAV splits and increases the headway. After the lane change of the LCV, the CAVs initially behind the LCV will be renumbered and join the platoon again.
    
    \begin{figure}
    	\centering
    	
    	\subcaptionbox{Cut-in scenario for lane change\label{fig:cutin scenario}}{\includegraphics[width=9cm]{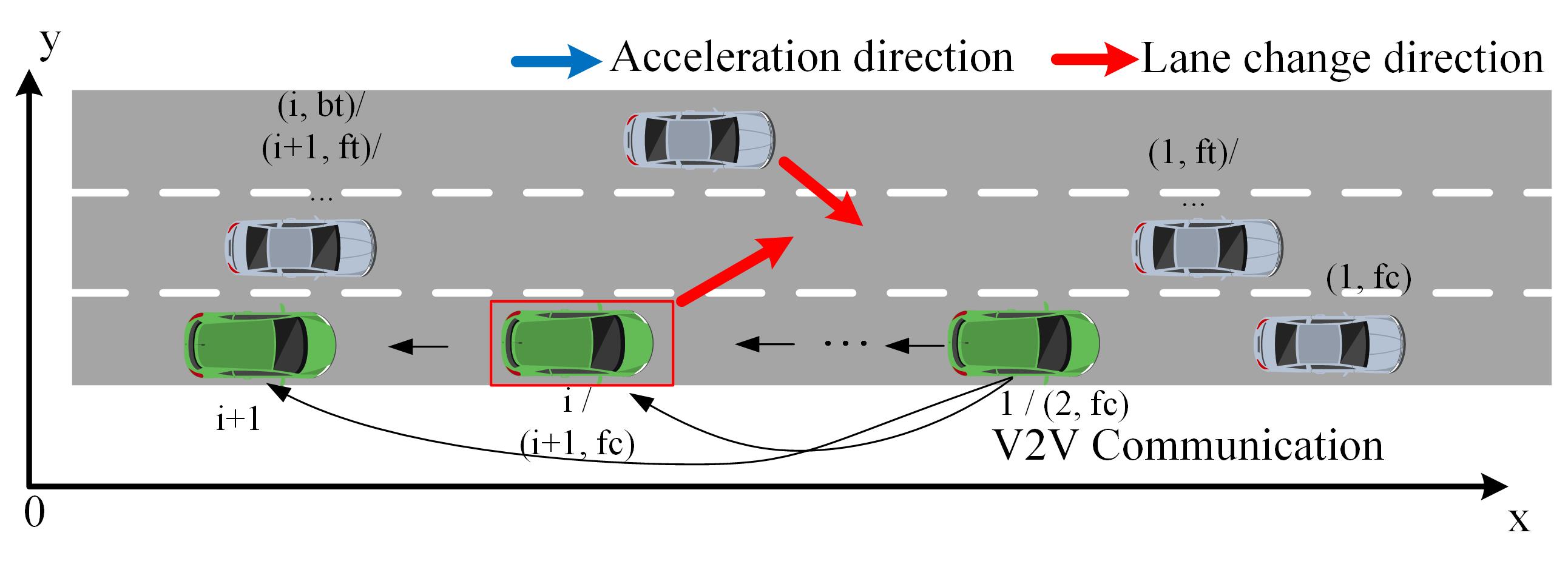}}\quad
    	\subcaptionbox{FDEC scenario for lane change\label{fig:fdec scenario}}{\includegraphics[width=9cm]{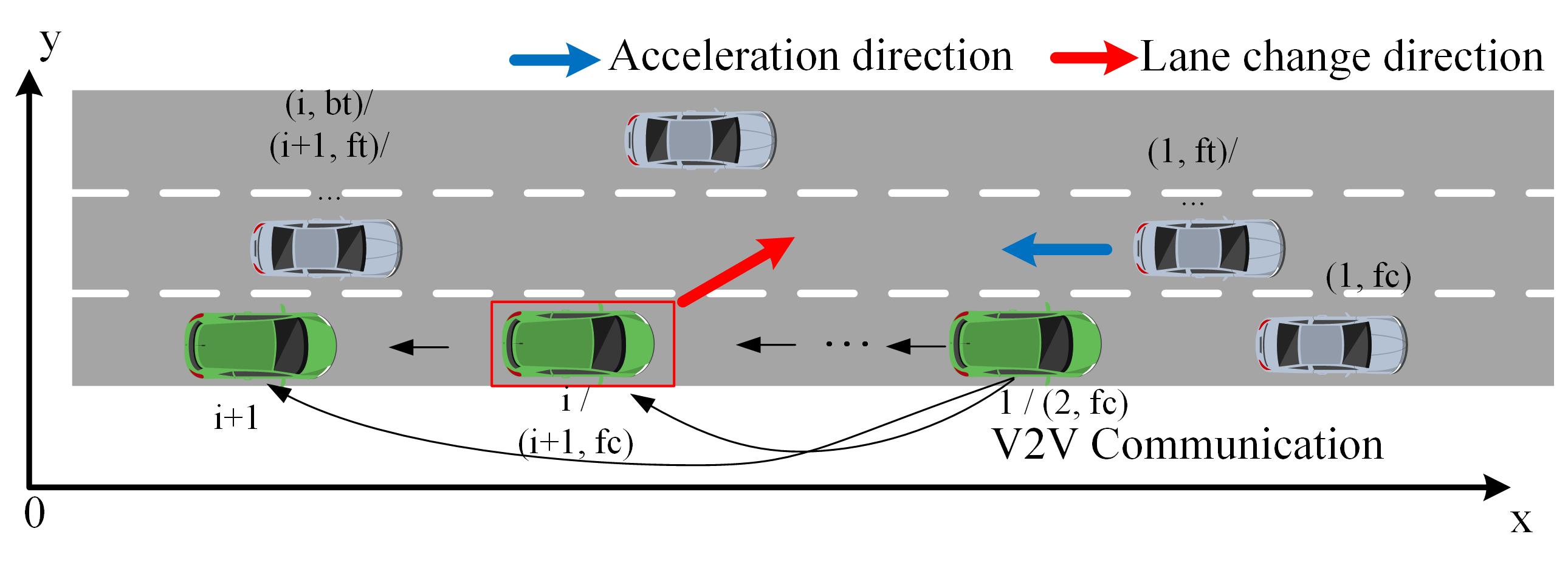}}\quad
    	\subcaptionbox{BACC scenario for lane change\label{fig:bacc scenario}}{\includegraphics[width=9cm]{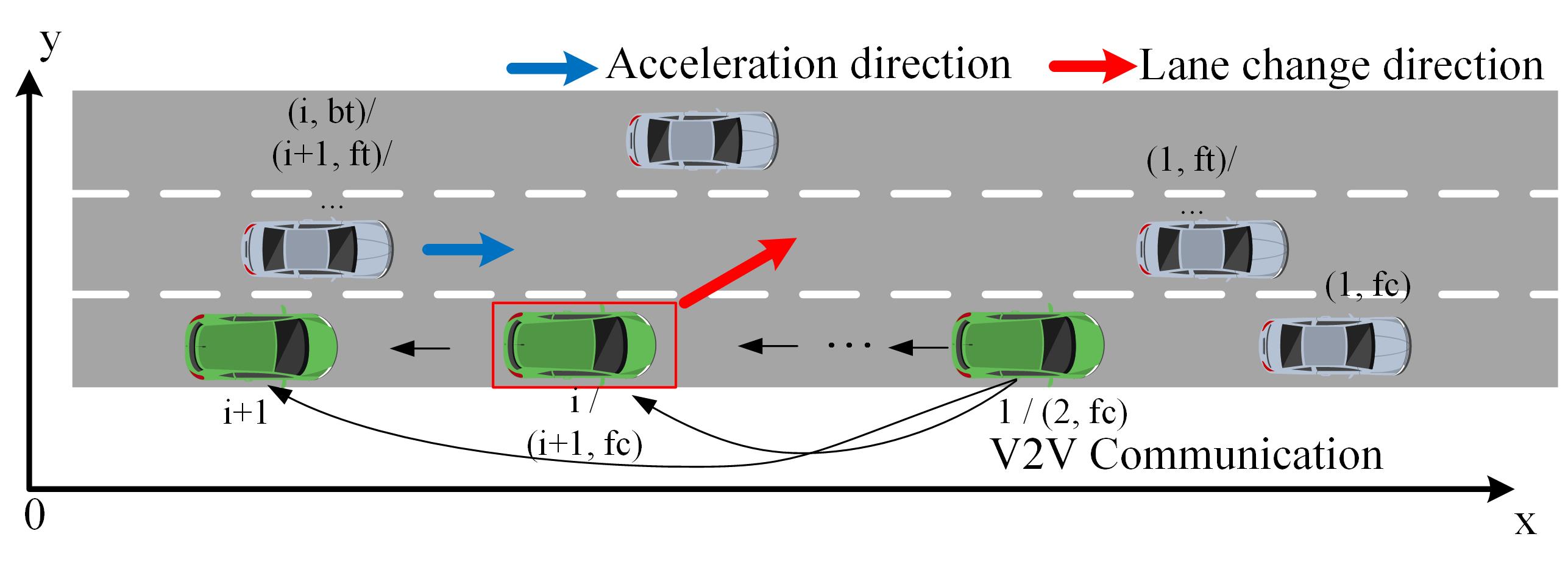}}\quad
    	\subcaptionbox{FFDEC scenario for lane change\label{fig:ffdec scenario}}{\includegraphics[width=9cm]{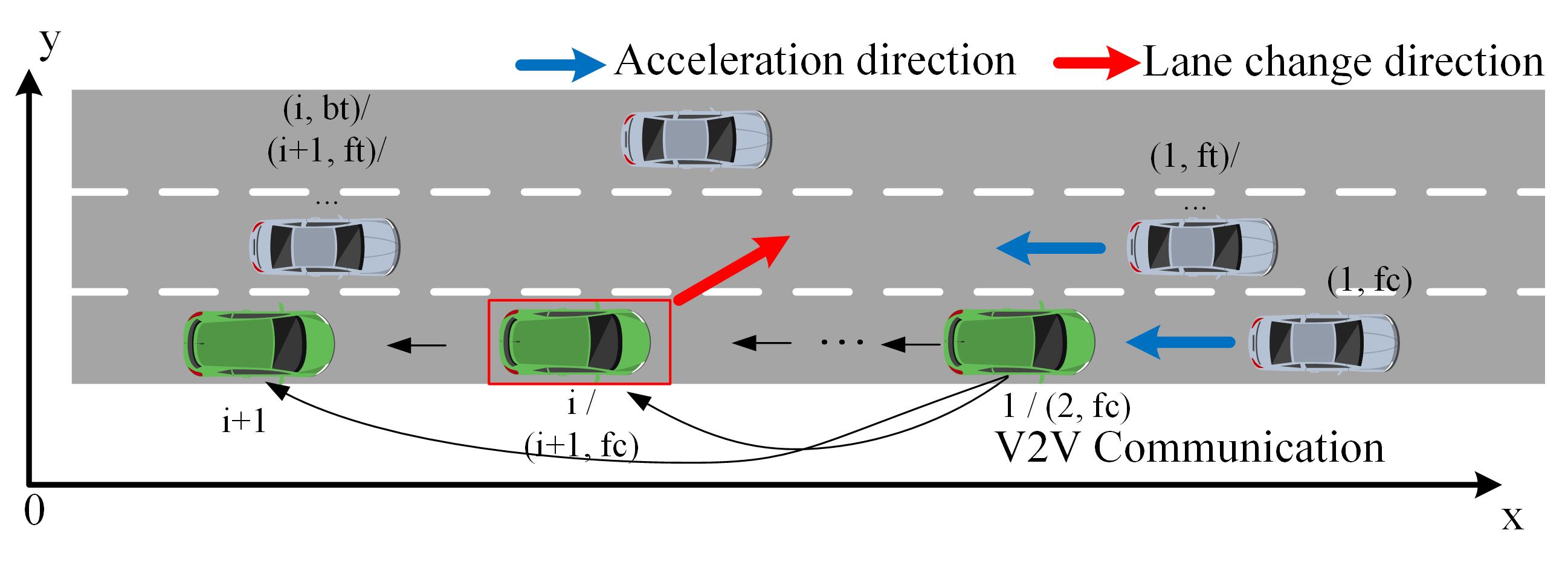}}
    	\caption{Safety-critical scenarios for lane change}\label{fig:Safety-critical scenarios for split}
    \end{figure}
    
    However, interacting with human-driven vehicles (HDVs), occasions are not rare when HDVs take an aggressive or even extreme maneuver that may cause collision. In this case, safety-critical control becomes crucial to both CAVs and the HDVs in the vicinity. These risky scenarios, in this paper, are entitled the safety-critical scenarios. In the literature, a widely used method for recognizing safety-critical scenarios is by calculating surrogate safety measures where time-to-collision (TTC) can be used as a safety indicator to determine whether the scenario could bring potential risk \cite{Vogel:2003}. Based on the TTC indicator, three categories of safety-critical scenarios that may cause very low TTC are considered for split operation:\\
    \textbf{Cut-in scenario:} lane change vehicle is cut in by other vehicles during lane change, shown in Fig.\ref{fig:cutin scenario}, where green vehicles are CAVs, and grey vehicles are HDVs.\\
    \textbf{FDEC scenario:} the front vehicle in the target lane decelerates when the CAV change its lane, shown in Fig.\ref{fig:fdec scenario}.\\
    \textbf{BACC scenario:} the back vehicle in the target lane accelerates when the CAV change its lane, shown in Fig.\ref{fig:bacc scenario}.\\
    \textbf{FFDEC scenario:} both the front vehicle in the target lane and the front vehicle in the current lane decelerates when the CAV change its lane, shown in Fig.\ref{fig:ffdec scenario}.
    
    In this problem, the basic assumptions are:
    
    \textbf{Assumption 1:} All the CAVs are equipped with sensors that enable the CAVs to extract necessary information from the environment, which include the position $(x, y)$ and the velocity $v$ of other vehicles. Considering the block phenomenon among vehicles in complicated traffic environments and the perception capabilities of the sensors, we assume that only the nearest vehicles in the current lane and the adjacent lanes can be sensed by each CAV. In this problem, for simplicity, only three of the sensed vehicles are considered for each CAV: the front vehicle in the current lane (fc), the front vehicle in the target lane (ft), the back vehicle in the target lane (bt). Based on this assumption, every considered HDV can be described by two indices: $(i, j)$, where $i\in \left\{ 1, 2, \cdots , n \right\} $ denotes that the vehicle is sensed by the $i$th CAV in the platoon; $j\in \left\{ \mathrm{fc}, \mathrm{ft}, \mathrm{bt} \right\}$ is defined as the relative position index of the sensed vehicle corresponding to the $i$th CAV. Note that a HDV can have multiple index pairs, but each index pair correspond to only one vehicle.
    
    \textbf{Assumption 2:} It is assumed that all the CAVs can communicate with each other by V2V communication. Four categories of perception information are shared between CAVs:
    \begin{itemize}
    	\item Vehicle position: $x$ and $y$ of the sensed vehicles.
    	\item Vehicle velocity: $v$ of the sensed vehicles.
    	\item Vehicle index pairs: $(i, j)$ of the sensed vehicles. $(i, 0)$ denotes the corresponding CAV, which can also be denoted as $i$ in the following parts for simplicity.
    	\item Command of Finite State Machine: $e, ps, pj, c, p$, only valid if the vehicle is a CAV.
    \end{itemize}

    \subsection{Control System Model of the CAV Platoon}
    In this paper, we use the simplified vehicle model in \cite{He:2021}. The states of the platoon system is shown as follow:
    \begin{equation}
    	\mathbf{x}=\left[ x_1, y_1, \psi _1, v_1, \cdots , x_n, y_n, \psi _n, v_n \right] ^{\mathrm{T}}
    \end{equation}
	where $i$ denotes the index of each CAV; $x_i$ and $y_i$ are the coordinates
	of the $i$th CAV's center of gravity; $\psi_i$ is heading angle of the $i$th CAV; $v_i$ is the velocity of the $i$th CAV. $l_r$ denotes the distance from the c.g. to the rear axles. In this paper, all the $l_r$ are assumed to be the same.
	The system inputs are defined as:
	\begin{equation}
		\mathbf{u}=\left[ a_1, \beta _1, \cdots , a_n, \beta _n \right] ^{\mathrm{T}}
	\end{equation}
	where $a_i$ denotes the acceleration of the $i$th CAV, and $\beta_i$ denotes the slip angle of the $i$th CAV.
	The platoon model is shown as below:
	\begin{equation}
		\mathbf{\dot{x}}=\boldsymbol{f}\left( \mathbf{x} \right) +\boldsymbol{g}\left( \mathbf{x} \right) \mathbf{u}
	\end{equation}
	This is an affine nonlinear control system where
	\begin{equation}
		\boldsymbol{f}\left( \mathbf{x} \right) =\left[ \mathbf{F}_{1}^{\mathrm{T}}, \mathbf{F}_{2}^{\mathrm{T}}, \cdots , \mathbf{F}_{n}^{\mathrm{T}} \right] ^{\mathrm{T}}
	\end{equation}

	\begin{equation}
		\mathbf{F}_i=\left[ v_i\cos \psi _i, v_i\sin \psi _i, 0, 0 \right] ^{\mathrm{T}}
	\end{equation}
	
	\begin{equation}
		\boldsymbol{g}\left( \mathbf{x} \right) =\left[ \mathbf{G}_{1}^{\mathrm{T}}, \mathbf{G}_{2}^{\mathrm{T}}, \cdots , \mathbf{G}_{n}^{\mathrm{T}} \right] ^{\mathrm{T}}
	\end{equation}
	
	\begin{equation}
		\mathbf{G}_i=\left[ \begin{matrix}
			\begin{array}{c}
				0\\
				-v_i\sin \psi _i\\
			\end{array}&		\begin{array}{c}
				0\\
				v_i\cos \psi _i\\
			\end{array}&		\begin{array}{c}
				0\\
				v_i/l_r\\
			\end{array}&		\begin{array}{c}
				1\\
				0\\
			\end{array}\\
		\end{matrix} \right] ^{\mathrm{T}}
	\end{equation}
	
	Given that the basic motivation of this study is to find a feedback control law for the lane change CAV to change lane while interact with the surrounding vehicles, the control law $\boldsymbol{q}(.)$ can be formulated as
	\begin{equation}
		\mathbf{u}=\boldsymbol{q}\left( \mathbf{x}, \mathbf{d} \right) 
	\end{equation}
	where $\mathbf{d}$ is the disturbance caused by the human-driven vehicles in the vicinity, defined by
	\begin{equation}
		\mathbf{d}=\left[ {\mathbf{d}_{1}^{\mathrm{fc}}}^{\mathrm{T}}, {\mathbf{d}_{1}^{\mathrm{ft}}}^{\mathrm{T}}, {\mathbf{d}_{1}^{\mathrm{bt}}}^{\mathrm{T}}, \cdots , {\mathbf{d}_{n}^{\mathrm{fc}}}^{\mathrm{T}}, {\mathbf{d}_{n}^{\mathrm{ft}}}^{\mathrm{T}}, {\mathbf{d}_{n}^{\mathrm{bt}}}^{\mathrm{T}} \right] ^{\mathrm{T}}
	\end{equation}
	
	\begin{equation}
		\mathbf{d}_{i}^{j}=\left[ x_{i}^{j}, y_{i}^{j}, \psi _{i}^{j} \right] ^{\mathrm{T}}
	\end{equation}
	where $i\in \left\{ 1, 2, \cdots , n \right\} $ denotes the index of a CAV in the platoon; $j\in \left\{ \mathrm{fc}, \mathrm{ft}, \mathrm{bt} \right\}$ is defined as the index of human-driven vehicles: fc - front vehicle in the current lane; ft - front vehicle in the target lane; bt - back vehicle in the target lane. For instance, $\mathbf{d}_{2}^{\mathrm{ft}}$ denotes the disturbance vector of the front human-driven vehicle in the target lane relative to the second CAV in the platoon.

    \subsection{Finite State Machine for Higher Level Controller}
	
	\begin{figure*}
		\centering
		\includegraphics[width=14cm]{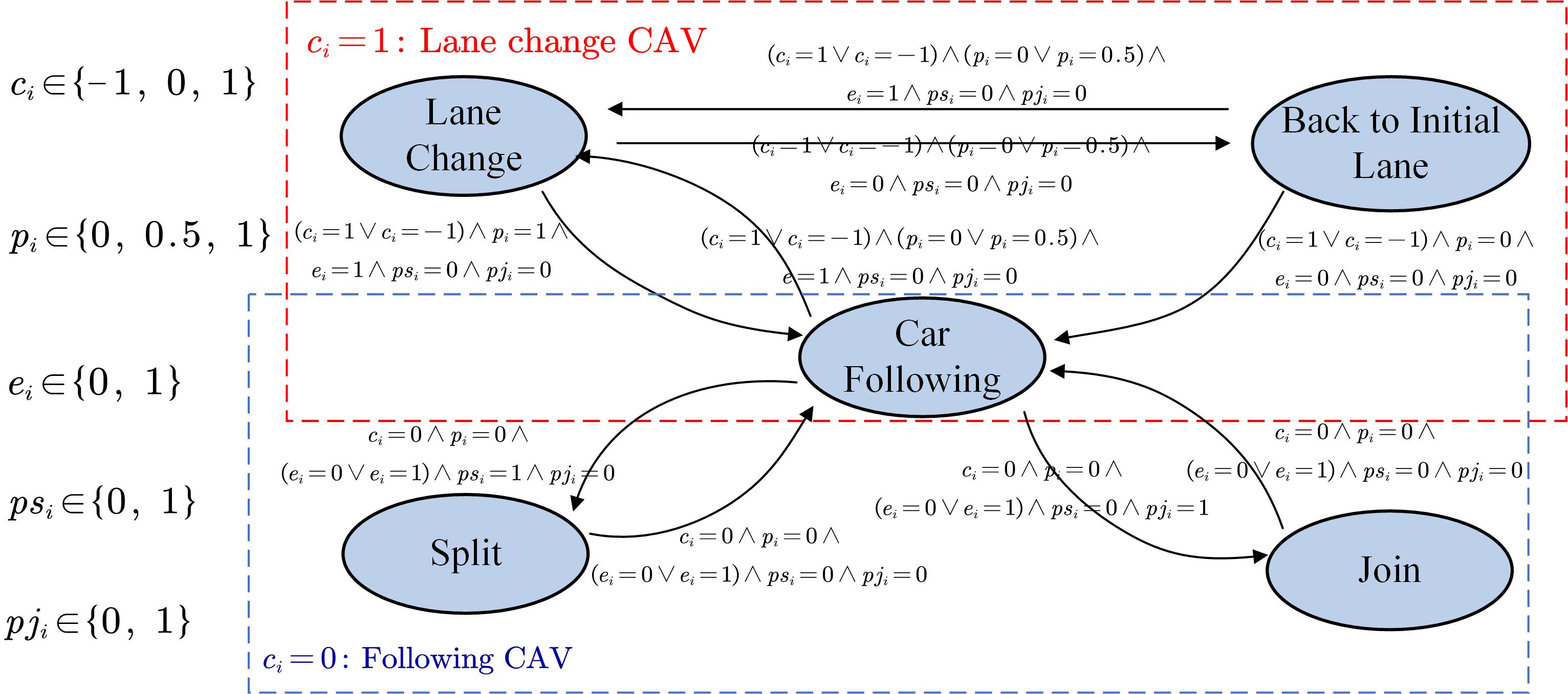}		
		\caption{Finite state machine for platoon operation control}\label{fig:FSM}
	\end{figure*}

    The two platoon operations are implemented by a two-layer lane change controller. The higher-level controller is a rule-based finite state machine (FSM). As shown in Fig.\ref{fig:FSM}, five states are considered: Car Following state, Lane Change state, Back to Initial Lane state, Split state, and Join state. These states are controlled by five signals:$e_i$, $ps_i$, $pj_i$, $c_i$, $p_i$, where $c_i=0$ means no lane change command, $c_i=1$ or $c_i=-1$ orders the CAV to change to the left or right lane respectively. $p_i=0$ means CAV in the original lane, $p_i=0.5$ means CAV is crossing the lane, $p_i=1$ means CAV have changed to the target lane. $e_i=1$ means safe lane change condition, $e_i=0$ means unsafe lane change condition, $ps_i=1$ means preparing for a split, $pj_i=1$ means preparing for a join. In the Car Following state, CAVs follow the preceding vehicle in a small headway, while in Split and Join state, CAVs will adjust the headway to allow changing lanes safely. The time headway profile in the split and the join state is shown in (\ref{eq:tau_split}), and the time headway profile in the Car Following state is shown in (\ref{eq:tau_join}).
    
    \begin{equation}
    	\tau _i =\left\{ \begin{array}{c}
    		0.6+0.2t\,\,  , if\,\, t<4\\
    		1.4             , if\,\, t\geqslant 4\\
    	\end{array} \right. 
    \label{eq:tau_split}
    \end{equation}

	\begin{equation}
		\tau _i =\left\{ \begin{array}{c}
			1.4-0.2t\,\,  , if\,\, t<4\\
			0.6             , if\,\, t\geqslant 4\\
		\end{array} \right.  
	\label{eq:tau_join}
	\end{equation}
	
	\subsection{CLF-CBF-QP for Lower Level Controller}
	To realize the lane change control, a lower level controller based on CLF-CBF-QP is designed for each CAV. For the $i$th CAV, the control input is decided by solving the following CLF-CBF-QP problem:
	\begin{equation}
		\begin{aligned}
			\mathbf{u}_{i}^{*}\,\,&=\,\,\underset{\left[ a_i\,\,\beta _i\,\,\delta _{l, i}\,\,\delta _{y, i}\,\,\delta _{\psi , i} \right]}{\mathrm{arg}\min}\,\,\frac{1}{2}{\mathbf{u}_i}^{\mathrm{T}}\mathbf{Hu}_i+p_l\delta _{l, i}^{2}+p_y\delta _{y, i}^{2}+p_{\psi}\delta _{\psi , i}^{2}\\
			\mathrm{s}.\mathrm{t}.&L_fV_{l, i}+L_gV_{l, i}\mathbf{u}_i\leqslant -\alpha _lV_{l, i}+\delta _{l, i}\\
			&\,\,L_fV_{y, i}+L_gV_{y, i}\mathbf{u}_i\leqslant -\alpha _yV_{y, i}+\delta _{y, i}\\
			&\,\,L_fV_{\psi , i}+L_gV_{\psi , i}\mathbf{u}_i\leqslant -\alpha _{\psi}V_{\psi , i}+\delta _{\psi , i}\\
			&\,\,\frac{\partial h_{i}^{j}}{\partial t}+L_fh_{i}^{j}+L_gh_{i}^{j}\mathbf{u}_i\geqslant -\gamma _jh_{i}^{j}\,\, ,  j\in \left\{ \mathrm{fc}, \mathrm{ft}, \mathrm{bt} \right\}\\
		\end{aligned}
	\end{equation}
	where $h_{i}^{j}$ represents the CBFs $h_{i}^{j}\left( \mathbf{x}_i, \mathbf{d}_{i}^{j},t \right) $. The first three constraints are CLF constraints, and the other three constraints concerning $h_{i}^{j}$ are CBF constraints.

	\subsection{Design of CLF Constraints}
	Different from the controller in \cite{He:2021}, to realize the split and join operation, the longitudinal Lyapunov candidate for the $i$th CAV is modified as
	\begin{equation}
		V_{l, i}=\alpha _1\left( v_i-v_{d, i} \right) ^2+\alpha _2\left( s_i-s_{d, i} \right) ^2
	\end{equation}
	where $s_i$ is the spacing gap defined by
	\begin{equation}
		\begin{aligned}
			&s_i=x_{i-2}-x_i-L \text{  for Join state}\\
			&s_i=x_{i-1}-x_i-L \text{  otherwise}
		\end{aligned}
	\end{equation}
	 $s_{d,i}$ is the desired spacing gap designed as $s_{d,i}=s_0+v_i\tau _i$, $\tau _ i$ is the desired time headway of the $i$th CAV, for $i>1$; $L$ is the vehicle length, $s_0$ is the desired spacing gap between CAVs when they do not move, $\alpha_1$ and $\alpha_2$ are parameters.
	The other two Lyapunov candidates are:
	\begin{equation}
		V_{y, i}=\left( y_y-y_{d, i} \right) ^2
	\end{equation}
	\begin{equation}
		V_{\psi ,i}={\psi _i}^2
	\end{equation}
	
	\subsection{Design of CBF Constraints}

	According to the previously defined indices, the CAVs in the platoon are numbered in a monotonically increasing order from head to tail. Each vehicle including CAVs and HDVs is specified by an index pair $(i, j)$, $i\in \left\{ 1, 2, \cdots , n \right\} $, $j\in \left\{ \mathrm{fc}, \mathrm{ft}, \mathrm{bt} \right\}$. For the $i$th CAV, control barrier functions $h_{i}^{(i, j)}$ can be designed to avoid a potential collision between vehicle $(i, j)$ and itself. For simplicity, $h_{i}^{j}$, $i\in \left\{ 1, 2, \cdots , n \right\} $, $j\in \left\{ \mathrm{fc}, \mathrm{ft}, \mathrm{bt} \right\}$ denotes CBFs between CAV $i$ and vehicle $(i, j)$; $h_{i_1}^{i_2}$, $i_1, i_2\in \left\{ 1, 2, \cdots , n \right\} $, denotes CBFs between different CAVs.
	
	\textbf{CBF in ACC state:} In ACC state, CAVs only need to consider the longitudinal collision, so the CBF is designed as (\ref{eq:cbf-acc-fc}) shows, where $a_{max}$ denotes the maximum acceleration of the vehicles, and $\epsilon_x$ denotes the longitudinal safety factor taking value from $0$ to $1$.
	\begin{figure*}
		\textbf{CBF in ACC state:}
		\centering
		\begin{equation}
			h_{i}^{fc}\left( \mathbf{x}_i,\mathbf{d}_{i}^{fc},t \right) =\begin{cases}
				v_{i}^{fc}t-x_i-\left( 1+\epsilon _x \right) v_i-\frac{\left( v_{i}^{fc}-v_i \right) ^2}{2a_{\max}}\,\,\mathrm{if}v_i\geqslant v_{i}^{fc}\\
				v_{i}^{fc}t-x_i-\left( 1+\epsilon _x \right) v_i\,\,\mathrm{else}\\
			\end{cases}
		\label{eq:cbf-acc-fc}
		\end{equation}
		
		\textbf{CBF in Lane Change state:}
		\centering
		\begin{equation}
			\begin{aligned}
				h_{i}^{fc}\left( \mathbf{x}_i,\mathbf{d}_{i}^{fc},t \right) =\begin{cases}
					v_{i}^{fc}t-x_i-\left( 1+\epsilon _x \right) v_i-\frac{\left( v_{i}^{fc}-v_i \right) ^2}{2a_{\max}}\,\,\mathrm{if}v_i\geqslant v_{i}^{fc}\\
					v_{i}^{fc}t-x_i-\left( 1+\epsilon _x \right) v_i\,\,\mathrm{else}\\
				\end{cases}
			\end{aligned}
			\label{eq:cbf-lc-fc}
		\end{equation}
		\begin{equation}
			\begin{aligned}
				h_{i}^{ft}\left( \mathbf{x}_i,\mathbf{d}_{i}^{ft}, t \right) =\begin{cases}
					v_{i}^{ft}t-x_i-\left( 1+\epsilon _x \right) v_i-\frac{\left( v_{i}^{ft}-v_i \right) ^2}{2a_{\max}}\,\,    \mathrm{if} v_i\geqslant v_{i}^{ft}\\
					v_{i}^{ft}t-x_i-\left( 1+\epsilon _x \right) v_i\,\,                     \mathrm{else}\\
				\end{cases}
			\end{aligned}
			\label{eq:cbf-lc-ft}
		\end{equation}
		\begin{equation}
			\begin{aligned}
				h_{i}^{bt}\left( \mathbf{x}_i,\mathbf{d}_{i}^{bt}, t \right) =\begin{cases}
					-v_{i}^{bt}t+x_i-\left( 1+\epsilon _x \right) v_{i}^{bt}-\frac{\left( v_{i}^{bt}-v_i \right) ^2}{2a_{\max}}\,\,    \mathrm{if} v_{i}^{bt}\geqslant v_i\\
					-v_{i}^{bt}t+x_i-\left( 1+\epsilon _x \right) v_{i}^{bt}\,\,                     \mathrm{else}\\
				\end{cases}
			\end{aligned}
			\label{eq:cbf-lc-bt}
		\end{equation}
		
		\textbf{CBF in Back to Current Lane state:}
		\centering
		\begin{equation}
			\begin{aligned}
				h_{i}^{fc}\left( \mathbf{x}_i,\mathbf{d}_{i}^{fc},t \right) =\begin{cases}
					v_{i}^{fc}t-x_i-\left( 1+\epsilon _x \right) v_i-\frac{\left( v_{i}^{fc}-v_i \right) ^2}{2a_{\max}}\,\,\mathrm{if}v_i\geqslant v_{i}^{fc}\\
					v_{i}^{fc}t-x_i-\left( 1+\epsilon _x \right) v_i\,\,\mathrm{else}\\
				\end{cases}
			\end{aligned}
		\label{eq:cbf-bcl-fc}
		\end{equation}
		\begin{equation}
			\begin{aligned}
				h_{i}^{ft}\left( \mathbf{x}_i,\mathbf{d}_{i}^{ft}, t \right) =\begin{cases}
					x_{i}^{ft}+v_{i}^{ft}t-x_i-\left( 1+\epsilon _x \right) v_i-\frac{\left( v_{i}^{ft}-v_i \right) ^2}{2a_{\max}}\,\,    \mathrm{if} x_{i}^{ft}-x_i\geqslant 0, v_i\geqslant v_{i}^{ft}\\
					x_{i}^{ft}+v_{i}^{ft}t-x_i-\left( 1+\epsilon _x \right) v_i\,\,                     \mathrm{if} x_{i}^{ft}-x_i\geqslant 0, v_i<v_{i}^{ft}\\
					y_{i}^{ft}-y_i-\epsilon _y\,\,                                      \mathrm{else}\\
				\end{cases}
			\end{aligned}
		\label{eq:cbf-bcl-ft}
		\end{equation}
		\begin{equation}
			\begin{aligned}
				h_{i}^{bt}\left( \mathbf{x}_i,\mathbf{d}_{i}^{bt}, t \right) =\begin{cases}
					-x_{i}^{bt}-v_{i}^{bt}t+x_i-\left( 1+\epsilon _x \right) v_{i}^{bt}-\frac{\left( v_{i}^{bt}-v_i \right) ^2}{2a_{\max}}\,\,    \mathrm{if} x_i-x_{i}^{bt}\geqslant 0, v_{i}^{bt}\geqslant v_i\\
					-x_{i}^{ft}-v_{i}^{ft}t+x_i-\left( 1+\epsilon _x \right) v_{i}^{bt}\,\,                     \mathrm{if} x_i-x_{i}^{bt}\geqslant 0, v_{i}^{bt}<v_i\\
					y_{i}^{bt}-y_i-\epsilon _y\,\,                                         \mathrm{else}\\
				\end{cases}
			\end{aligned}
		\label{eq:cbf-bcl-bt}
		\end{equation}
		
		\textbf{CBF in Split state:}
		\centering
		\begin{equation}
			\begin{aligned}
				h_{i}^{i-1}\left( \mathbf{x}_i,\mathbf{x}_{i-1}, t \right) =\begin{cases}
					x_{i-1}+v_{i-1}t-x_i-\left( 1+\epsilon _x \right) v_i-\frac{\left( v_{i-1}-v_i \right) ^2}{2a_{\max}}\,\,    \mathrm{if} v_i\geqslant v_{i-1}\\
					x_{i-1}+v_{i-1}t-x_i-\left( 1+\epsilon _x \right) v_i\,\,                       \mathrm{else}\\
				\end{cases}
			\end{aligned}
		\label{eq:cbf-split}
		\end{equation}	
		
		\textbf{CBF in Join state:}
		\centering
		\begin{equation}
			\begin{aligned}
				h_{i}^{i-2}\left( \mathbf{x}_i,\mathbf{x}_{i-2}, t \right) =\begin{cases}
					x_{i-2}+v_{i-2}t-x_i-\left( 1+\epsilon _x \right) v_i-\frac{\left( v_{i-2}-v_i \right) ^2}{2a_{\max}}\,\,    \mathrm{if} v_i\geqslant v_{i-2}\\
					x_{i-2}+v_{i-2}t-x_i-\left( 1+\epsilon _x \right) v_i\,\,                        \mathrm{else}\\
				\end{cases}
			\end{aligned}
		\label{eq:cbf-join}	
		\end{equation}	
	\end{figure*}

	\textbf{CBF in Lane Change state:} In Lane Change state, the lane-change CAV should avoid the lateral collision, so the CBF is designed as (\ref{eq:cbf-lc-fc})(\ref{eq:cbf-lc-ft})(\ref{eq:cbf-lc-bt}) show.

	\textbf{CBF in Back to Current Lane state:} In Back to Current Lane state, the lane-change CAV should also avoid the lateral collision. However, when CAV is back to current lane, it is fine to overlap with vehicles in the target lane longitudinally, only if they do not laterally collide. Thus, the CBF is designed as (\ref{eq:cbf-bcl-fc})(\ref{eq:cbf-bcl-ft})(\ref{eq:cbf-bcl-bt}) show, where $\epsilon_ y$ is a small lateral safety factor.

	\textbf{CBF in Split state:} In this case, the vehicle just behind the lane change CAV (CAV $i-1$) is in Split state. It tries to make room for the lane change CAV in case of any safety-critical scenarios where the lane-change CAV may drive back to the current lane and rejoin the platoon. So, the CBF in this state is designed to keep a safe longitudinal distance with the lane-change CAV even if it is not running in the current lane, as shown in (\ref{eq:cbf-split}).

	\textbf{CBF in Join state:} In this case, after the lane-change CAV finishes the lane change, the headway of the CAVs will decrease gradually. In the Join state, the CBF is designed to keep a safe longitudinal distance with the front CAV (CAV $i-2$) in the platoon, as shown in (\ref{eq:cbf-join}).

	\section{Experiment}
	Four comparative experiments are done in the corresponding safety-critical scenarios. In these comparative experiments, the parameters listed in Table \ref{tab:universal parameters} are not changed. The initial states of the CAVs in the first three scenarios are presented in Table \ref{tab:initial states of CAVs}.
	
		\begin{table}[htbp]
		\caption{Universal Parameters in All the experiments}
		\begin{center}
			\begin{tabular}{|c|c|c|c|}
				\hline
				\textbf{Parameter} & \textbf{Value}& \textbf{Parameter}& \textbf{Value} \\
				\hline
				$l_r$& $1.74$ & $l_f$ & $1.11$ \\
				\hline
				$l_{fc}$& $2.15$ & $l_{rc}$ & $2.77$ \\
				\hline
				$w$& $1.86$ & $dt$ & $0.05$ \\
				\hline
			\end{tabular}
			\label{tab:universal parameters}
		\end{center}
	\end{table}
	
	\begin{table}[htbp]
		\caption{The Initial States of the CAVs}
		\begin{center}
			\begin{tabular}{|c|c|c|c|c|}
				\hline
				& $x$& $y$ & $\psi$ & $v$ \\
				\hline
				CAV1& $100$ & $1.8$ & $0$ & $27.5$  \\
				\hline
				CAV2& $50$ & $1.8$ & $0$ & $27.5$\\
				\hline
				CAV3& $0$ & $1.8$ & $0$ & $27.5$ \\
				\hline
			\end{tabular}
			\label{tab:initial states of CAVs}
		\end{center}
	\end{table}
	
 	\subsection{Compare CBF-CLF-QP and CLF-QP in Cut-in Scenario}
 
 	\begin{figure*}
 		\centering
 		
 		\subcaptionbox{Collision-related states of CLF-CBF-QP-based experiment \label{fig:cutin-cbf-collision}}{\includegraphics[width=9cm]{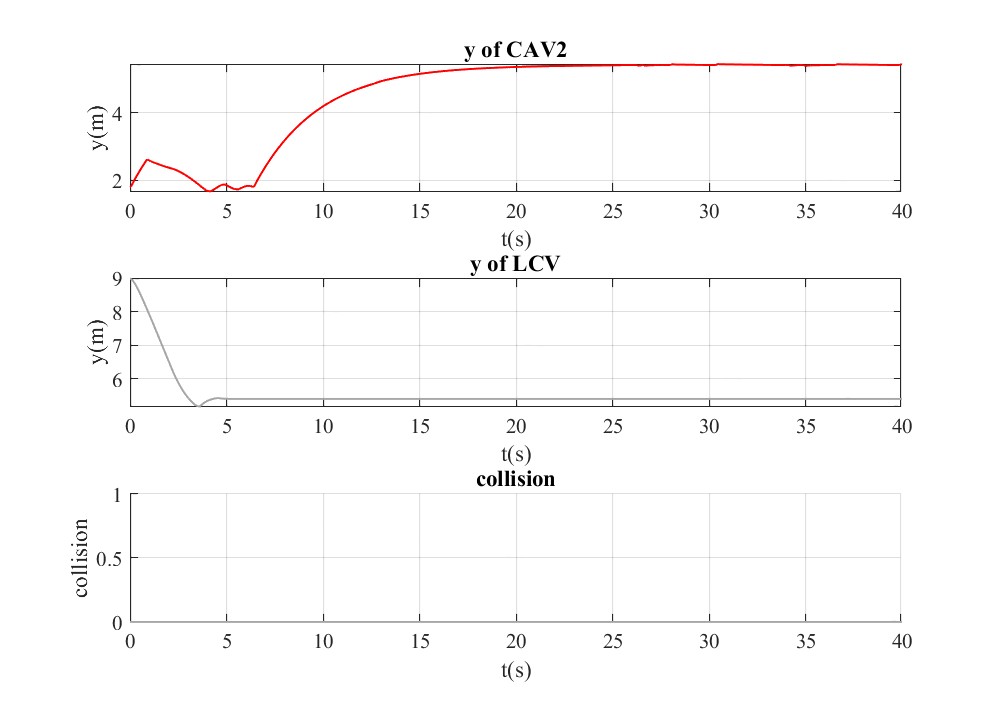}}
 		\subcaptionbox{Velocities and trajectory of CLF-CBF-QP-based experiment \label{fig:cutin-cbf-v-traj}}{\includegraphics[width=9cm]{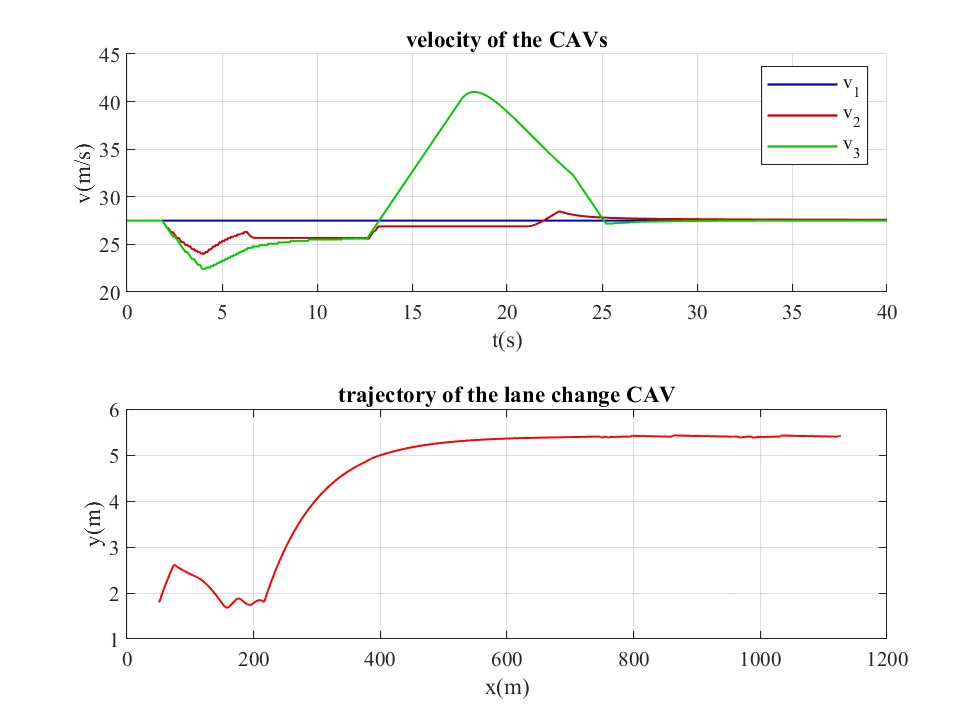}}
 		\quad
 		
 		\subcaptionbox{Collision-related states of CLF-QP-based experiment \label{fig:cutin-nocbf-collision}}{\includegraphics[width=9cm]{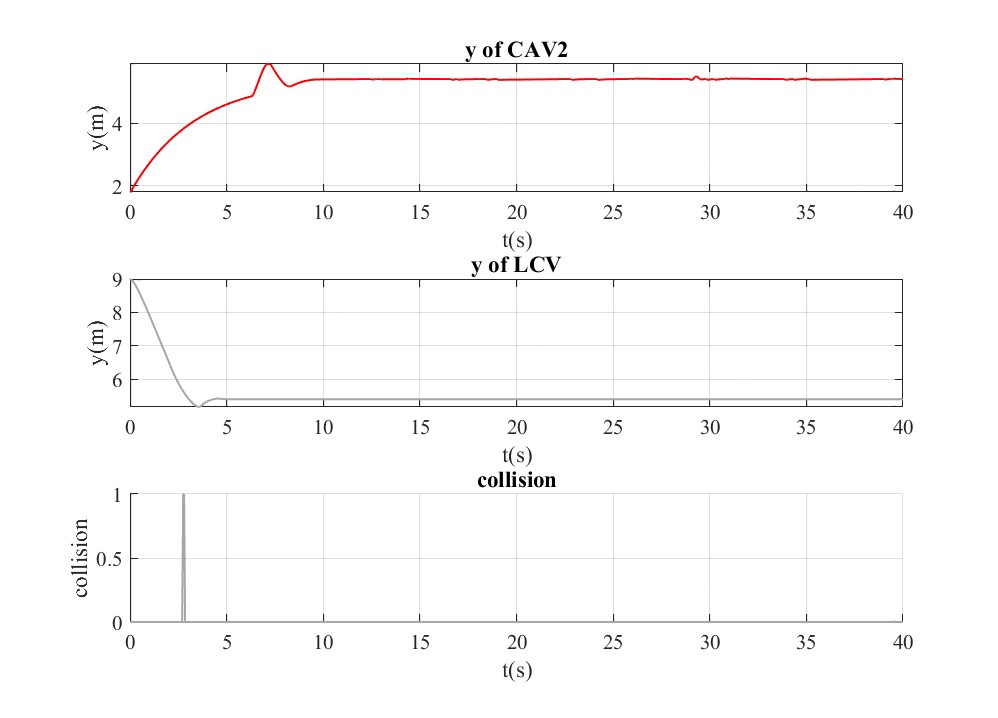}}
 		\subcaptionbox{Velocities and trajectory of CLF-QP-based experiment \label{fig:cutin-nocbf-v-traj}}{\includegraphics[width=9cm]{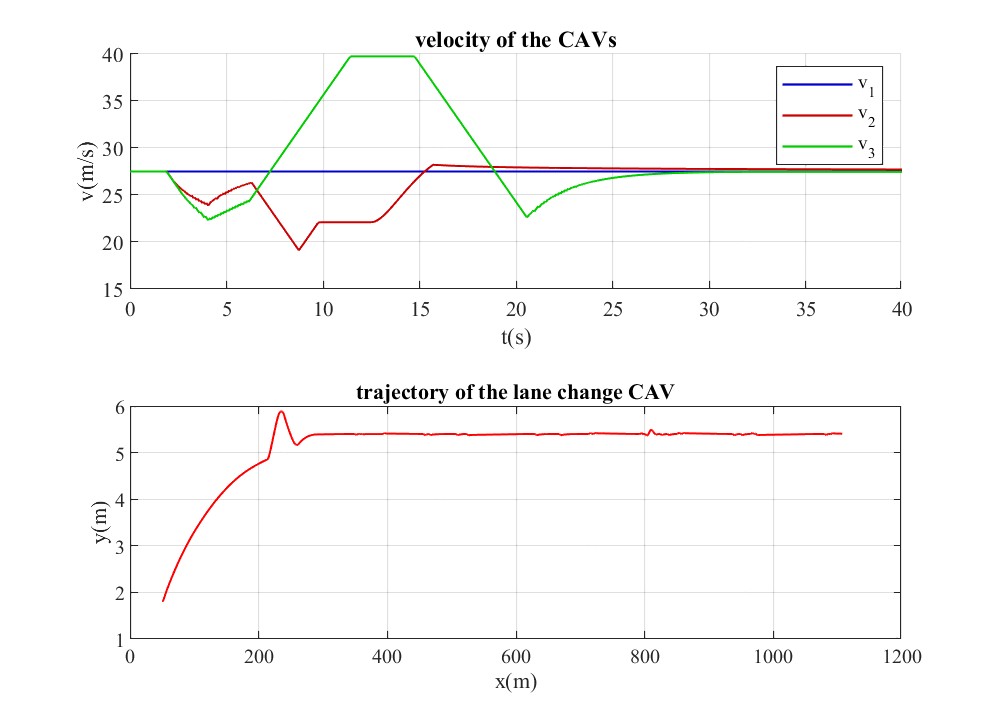}}
 		
 		\caption{Result of comparative experiment in Cut-in scenario}\label{fig:result-cutin-cbf}
 	\end{figure*}
 
 	To show the effectiveness of the designed CBF-CLF-QP controller, two comparative experiments are done in cut-in scenario using numerical simulation. In this experiment, three CAVs are running in a platoon in the bottom lane of the road, of which the lane width is $3.6\text{m}$. The second CAV is trying to change to the middle lane while a human-driven vehicle driving in the top lane, denoted as LCV, also tries to change to the middle lane. In this safety-critical scenario, these four vehicles are important. The initial states of the LCV are $x=45\text{m}$, $y=9\text{m}$, $\psi=0\text{rad}$, $v=27\text{m/s}$. The LCV starts at the top lane, and conducted a cut-in maneuver in ignorant of any danger.
 	
 	The controller parameters are: $\alpha_l=1.7$, $\alpha_y=0.6$, $\alpha_{\psi}=18$, $p_l=15$, $p_y=0.05$, $p_{\psi}=400$. The result of CLF-CBF-QP controller is shown in Fig.\ref{fig:result-cutin-cbf}. According to Fig.\ref{fig:cutin-cbf-collision} and Fig.\ref{fig:cutin-cbf-v-traj}, when the LCV tried to compete for the target lane, the CBF constraints of the lane-change CAV could not be satisfied, so that the lane-change CAV switched to Back to Lane state and took an evasive action. The binary collision indicator shows $1$ if there is a collision and shows 0 if not. The lane-change CAV did not switched to Lane Change state again until the CBF constraints were then satisfied after the LCV finished lane change. Finally, the lane change CAV successfully changed to the target lane, and the platoon of other CAVs joined again after the split.
 	
 	The result of CLF-QP controller is shown in Fig.\ref{fig:cutin-nocbf-collision} and Fig.\ref{fig:cutin-nocbf-v-traj}, the LCV and the lane-change CAV collided during the lane change.
 	
 	\subsection{Compare CBF-CLF-QP and CLF-QP in FDEC Scenario}
 	
 	Similarly, comparative experiments by  numerical simulation are conducted under FDEC scenario. In this experiment, the second CAV is trying to change to the middle lane while a human-driven vehicle driving in the target lane in the front, denoted as FTV, took an abrupt deceleration. In this safety-critical scenario, these four vehicles are important. The initial states of the FTV are $x=75\text{m}$, $y=5.4\text{m}$, $\psi=0\text{rad}$, $v=30\text{m/s}$. The FTV keeps the initial speed until $t=1.8\text{s}$ when the FTV brakes with a deceleration of $-9\rm{m/s^2}$. The FTV accelerates again at $t=3\text{s}$ before the speed reaches $31\text{m/s}$ and then follow this speed until the experiment ends.
 	
 	\begin{figure*}
 		\centering
 		
 		\subcaptionbox{Collision-related states of CLF-CBF-QP-based experiment \label{fig:fdec-cbf-collision}}{\includegraphics[width=9cm]{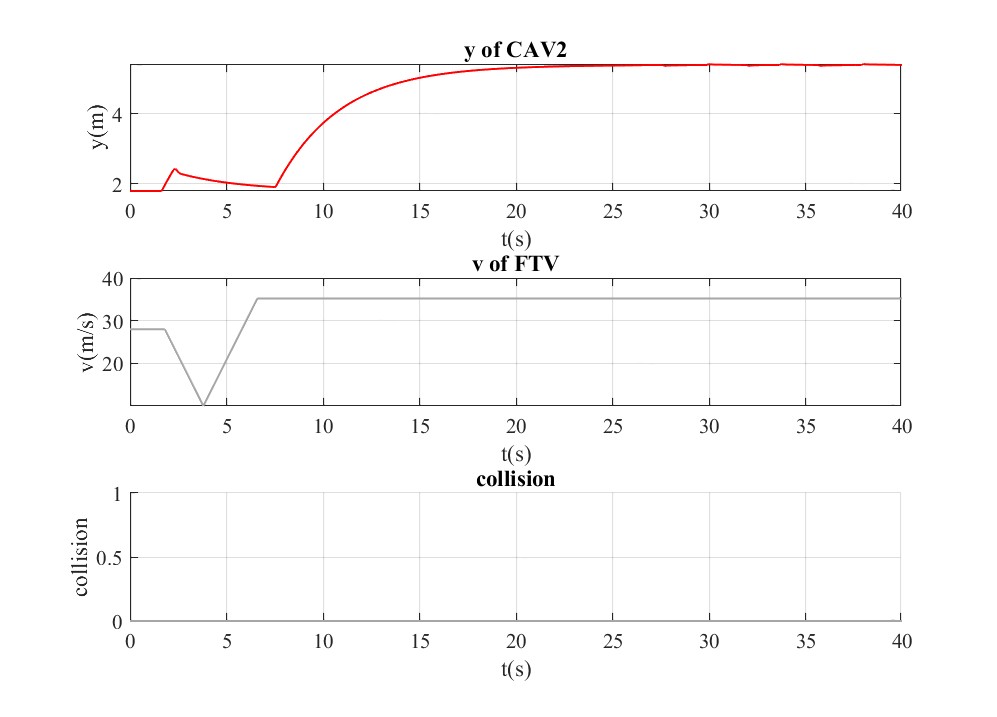}}
 		\subcaptionbox{Velocities and trajectory of CLF-CBF-QP-based experiment \label{fig:fdec-cbf-v-traj}}{\includegraphics[width=9cm]{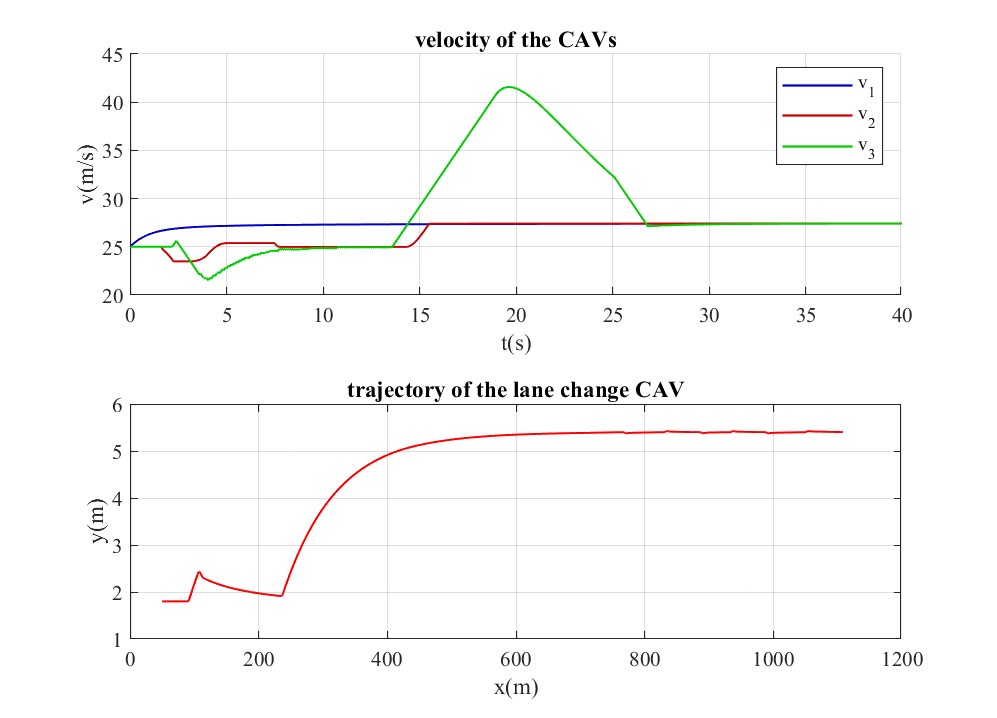}}
 		\quad
 		
 		\subcaptionbox{Collision-related states of CLF-QP-based experiment \label{fig:fdec-nocbf-collision}}{\includegraphics[width=9cm]{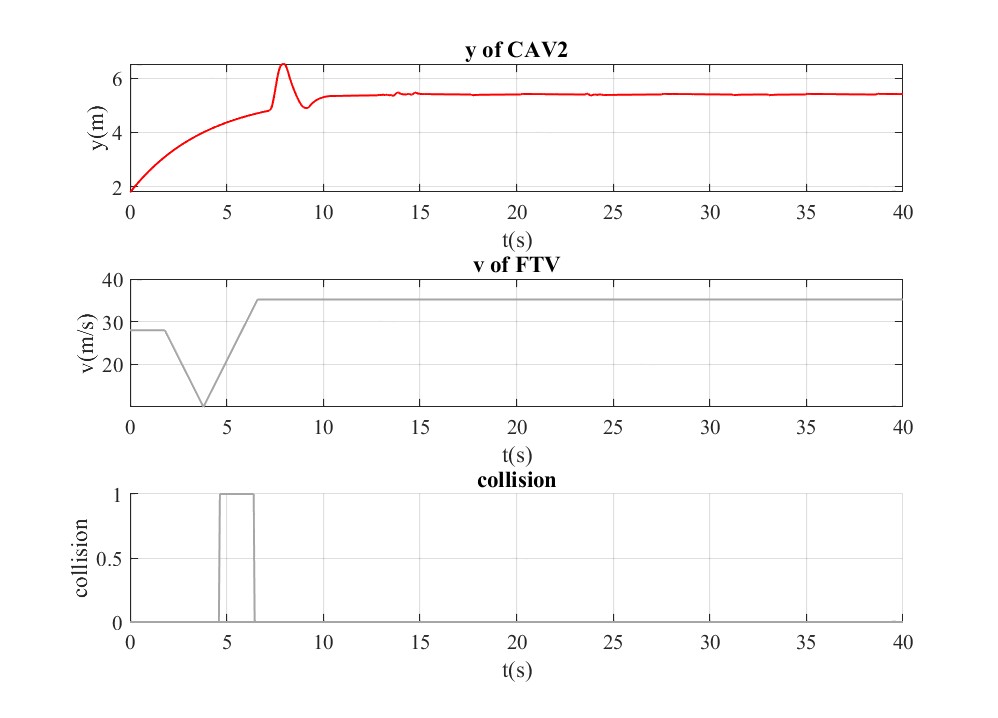}}
 		\subcaptionbox{Velocities and trajectory of CLF-QP-based experiment \label{fig:fdec-nocbf-v-traj}}{\includegraphics[width=9cm]{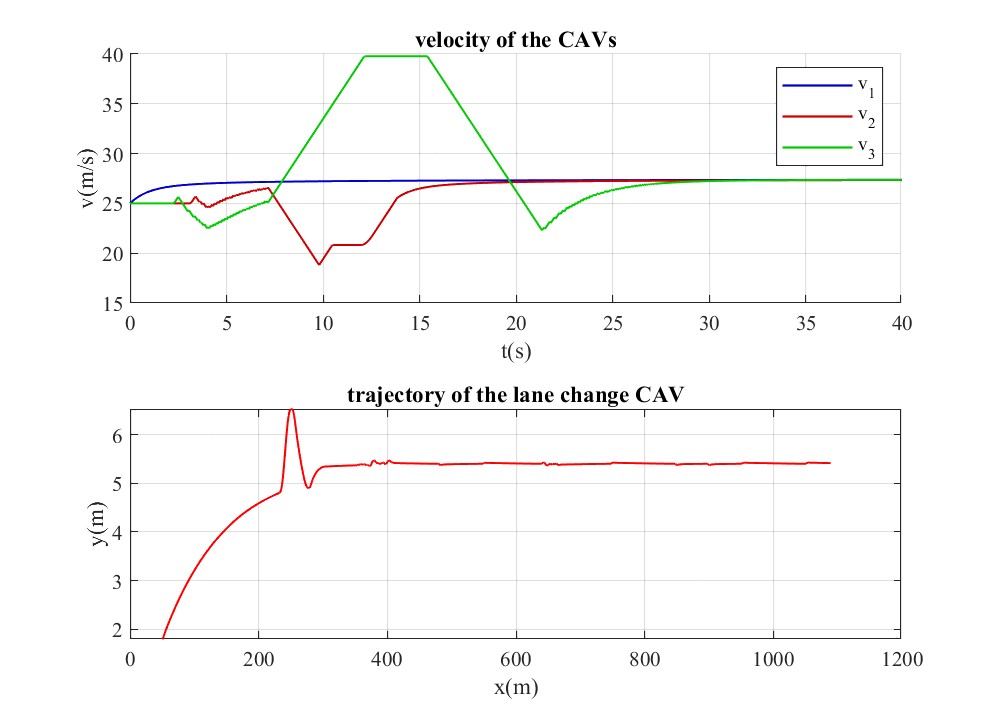}}
 		
 		\caption{Result of comparative experiment in FDEC scenario}\label{fig:result-fdec-cbf}
 	\end{figure*}
 	
 	The controller parameters are: $\alpha_l=1.7$, $\alpha_y=0.6$, $\alpha_{\psi}=20$, $p_l=15$, $p_y=0.02$, $p_{\psi}=400$. The result of CLF-CBF-QP controller is shown in Fig.\ref{fig:result-fdec-cbf}. According to Fig.\ref{fig:fdec-cbf-collision} and Fig.\ref{fig:fdec-cbf-v-traj}, as the FTV suddenly braked, the CBF constraints of the lane-change CAV worked, so that the lane-change CAV switched to Back to Lane state. After the FTV drove fast and far away enough, the scenario was not risky anymore. The lane-change CAV tried again and finally changed to the target lane.
 	
 	The result of CLF-QP controller is shown in Fig.\ref{fig:fdec-nocbf-collision} and Fig.\ref{fig:fdec-nocbf-v-traj} where the FTV and the lane-change CAV collided during the lane change. It is apparent that the lane-change CAV decelerated when the FTV braked. This is because the designed longitudinal CLF tries to keep a desired distance between vehicles. Although the proposed CLF is capable of collision prevention to some extent, it cannot cope with some extreme safety-critical scenarios such as the scenario in this experiment. In this case, the CBF is necessary to stay safe when doing lane change.
 	
 	\subsection{Compare CBF-CLF-QP and CLF-QP in BACC Scenario}
 	
 	Two other numerical simulation experiments are conducted under BACC scenario. In these experiments, the three CAVs are controlled by the same controller as those in FDEC scenario. The BTV denotes the back vehicle in the target lane. The initial states of the BTV are $x=0\text{m}$, $y=5.4\text{m}$, $\psi=0\text{rad}$, $v=30\text{m/s}$.  Similarly, the BTV keeps $30\text{m/s}$ for $t=1.8\text{s}$ and accelerates with an acceleration of $9\rm{m/s^2}$. It decelerates at $t=3\text{s}$ with an deceleration of $-3\rm{m/s^2}$ to $22\text{m/s}$ and then follow this speed until the experiment ends.

 	\begin{figure*}
 		\centering
 		
 		\subcaptionbox{Collision-related states of CLF-CBF-QP-based experiment \label{fig:bacc-cbf-collision}}{\includegraphics[width=9cm]{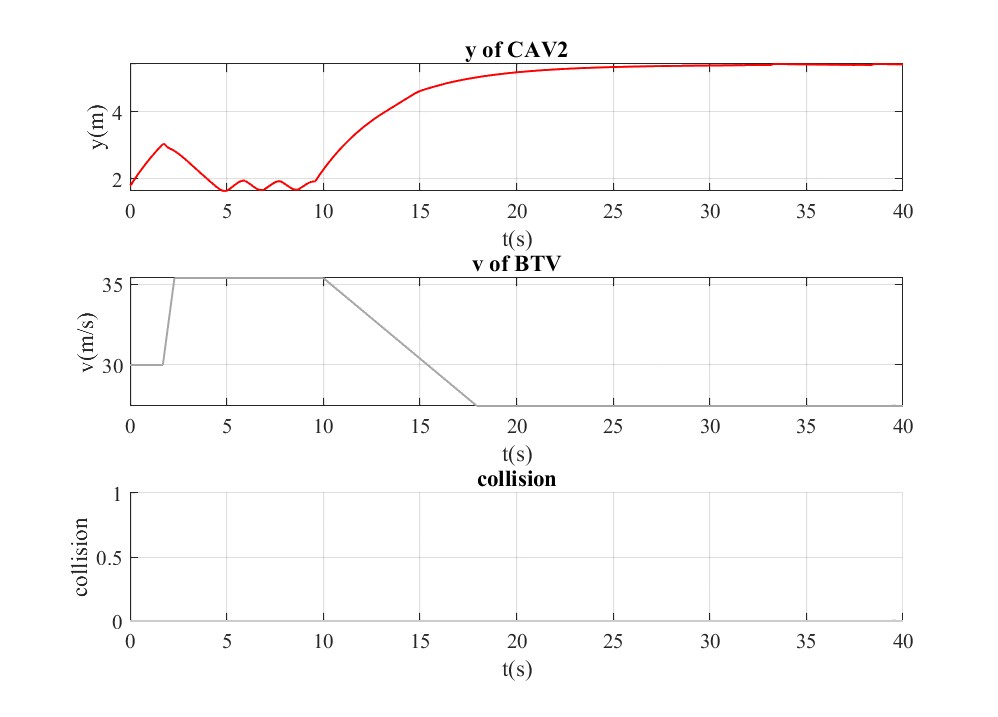}}
 		\subcaptionbox{Velocities and trajectory of CLF-CBF-QP-based experiment \label{fig:bacc-cbf-v-traj}}{\includegraphics[width=9cm]{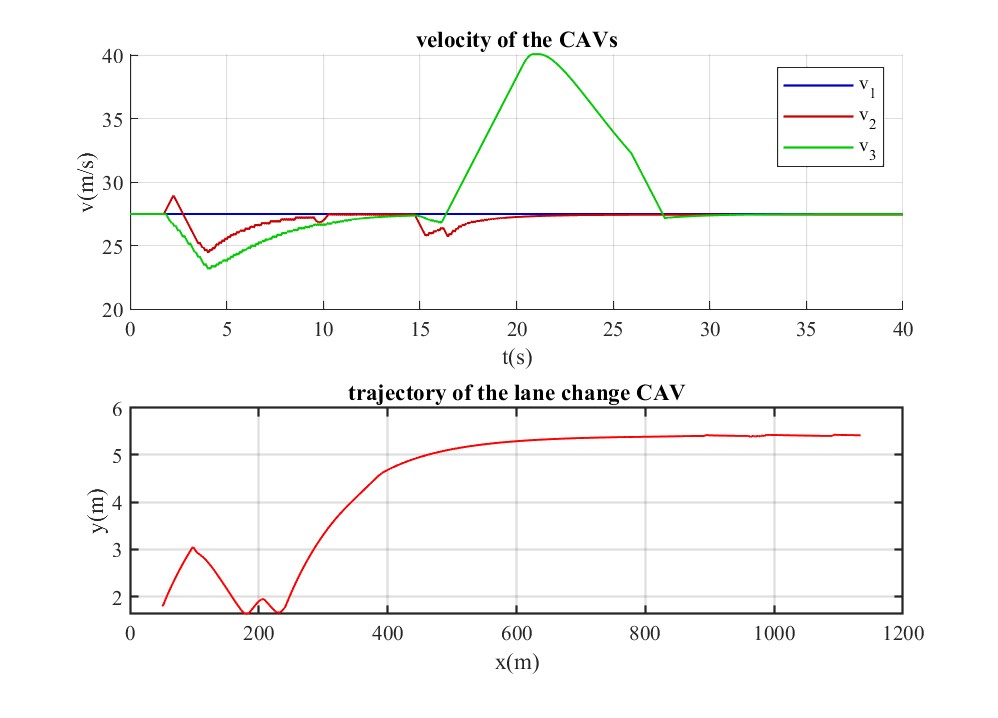}}
 		\quad
 		
 		\subcaptionbox{Collision-related states of CLF-QP-based experiment \label{fig:bacc-nocbf-collision}}{\includegraphics[width=9cm]{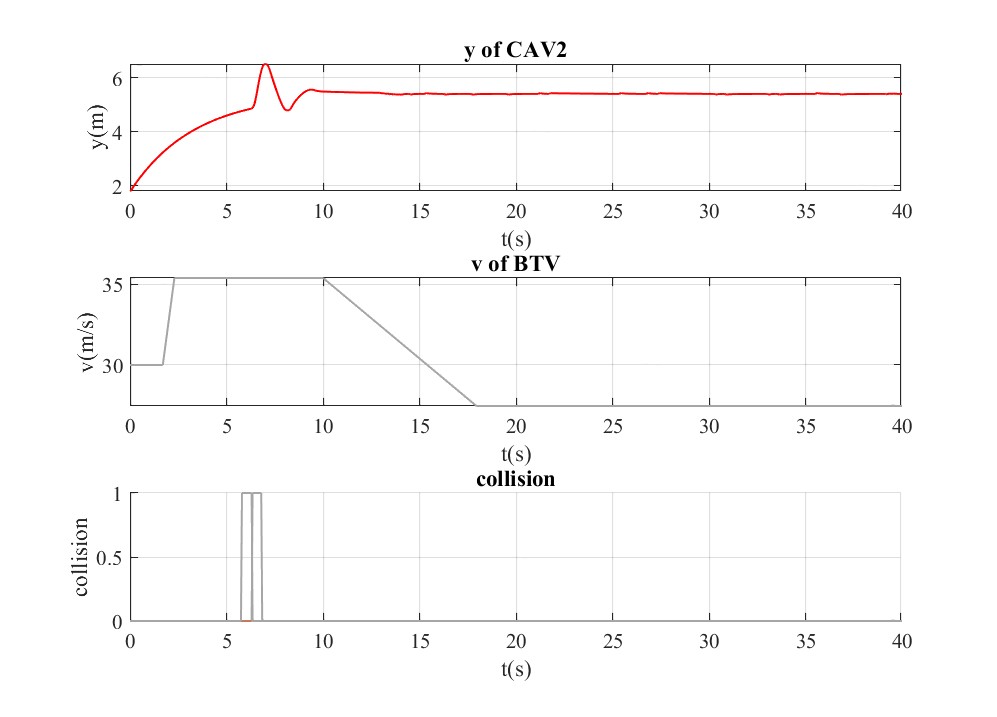}}
 		\subcaptionbox{Velocities and trajectory of CLF-QP-based experiment \label{fig:bacc-nocbf-v-traj}}{\includegraphics[width=9cm]{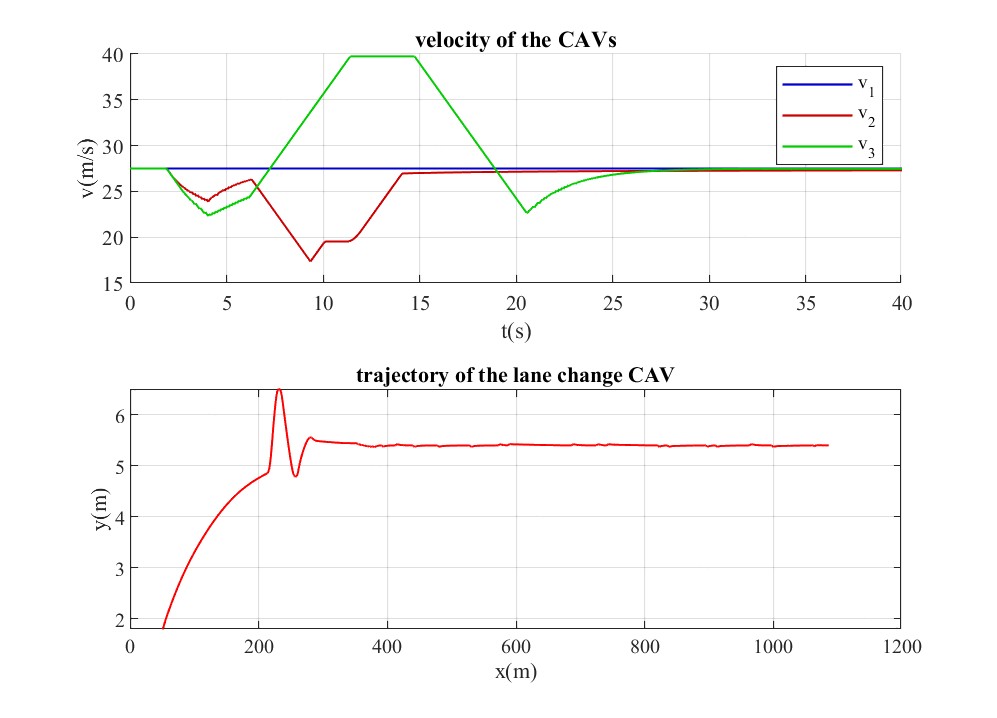}}
 		
 		\caption{Result of comparative experiment in BACC scenario}\label{fig:result-bacc-cbf}
 	\end{figure*}

 	The controller parameters are: $\alpha_l=1.7$, $\alpha_y=0.6$, $\alpha_{\psi}=18$, $p_l=15$, $p_y=0.02$, $p_{\psi}=900$. The result of the CLF-CBF-QP controller is shown in Fig.\ref{fig:bacc-cbf-collision} and Fig.\ref{fig:bacc-cbf-v-traj}. as the BTV suddenly accelerated, the lane-change CAV switched to Back to Lane state and returned to the initial lane until the BTV decelerated. Finally, the lane change CAV successfully changed to the target lane without any collision, and the platoon of other CAVs joined again.
 	
 	The result of CLF-QP controller is shown in Fig.\ref{fig:bacc-nocbf-collision} and Fig.\ref{fig:bacc-nocbf-v-traj}, the BTV and the lane-change CAV collided during the lane change. As is shown in Fig.\ref{fig:bacc-nocbf-v-traj}, there is a strong oscillation after the collision. This is because the lane-change CAV was too close to the BTV. The CLF-QP problem 'focused' on satisfying the longitudinal CLF constraints, so that the lateral CLF constraints did not contribute much to solving the control input. 
 	
 	\subsection{Compare CBF-CLF-QP for platoon operation and CBF-CLF-QP for single vehicle in FFDEC Scenario}
 	\begin{table}[htbp]
 		\caption{The Initial States in FFDEC Experiment}
 		\begin{center}
 			\begin{tabular}{|c|c|c|c|c|}
 				\hline
 				& $x$& $y$ & $\psi$ & $v$ \\
 				\hline
 				CAV1& $60$ & $1.8$ & $0$ & $20$  \\
 				\hline
 				CAV2& $30$ & $1.8$ & $0$ & $20$\\
 				\hline
 				CAV3& $0$ & $1.8$ & $0$ & $20$ \\
 				\hline
 				FTV& $70$ & $5.4$ & $0$ & $20$ \\
 				\hline
 				FCV& $90$ & $1.8$ & $0$ & $20$ \\
 				\hline
 			\end{tabular}
 			\label{tab:initial states of FFDEC exp}
 		\end{center}
 	\end{table}
 
 	Two numerical simulation experiments are conducted in FFDEC scenario to demonstrate the effectiveness of platoon operation. In these experiments, the FTV denotes the front vehicle in the target lane. The FCV denotes the front vehicle in the current lane. These five vehicle's initial states are presented in Table \ref{tab:initial states of FFDEC exp}.  The FTV keeps $20\text{m/s}$ for $t=1\text{s}$ and decelerates with an deceleration of $-6\rm{m/s^2}$. It accelerates at $t=3.8\text{s}$ with an acceleration of $3\rm{m/s^2}$ to $28\text{m/s}$ and then follow this speed until the experiment ends. The FCV follows a similar speed profile, but it decelerates to $15\text{m/s}$ at $t=1\text{s}$ with an deceleration of $-1\rm{m/s^2}$ during the deceleration period.
 	
\begin{figure*}
 		\centering
 		
 		\subcaptionbox{Selected states of key vehicles in cooperation-based experiment \label{fig:fdec2-withcoop-collision}}{\includegraphics[width=9cm]{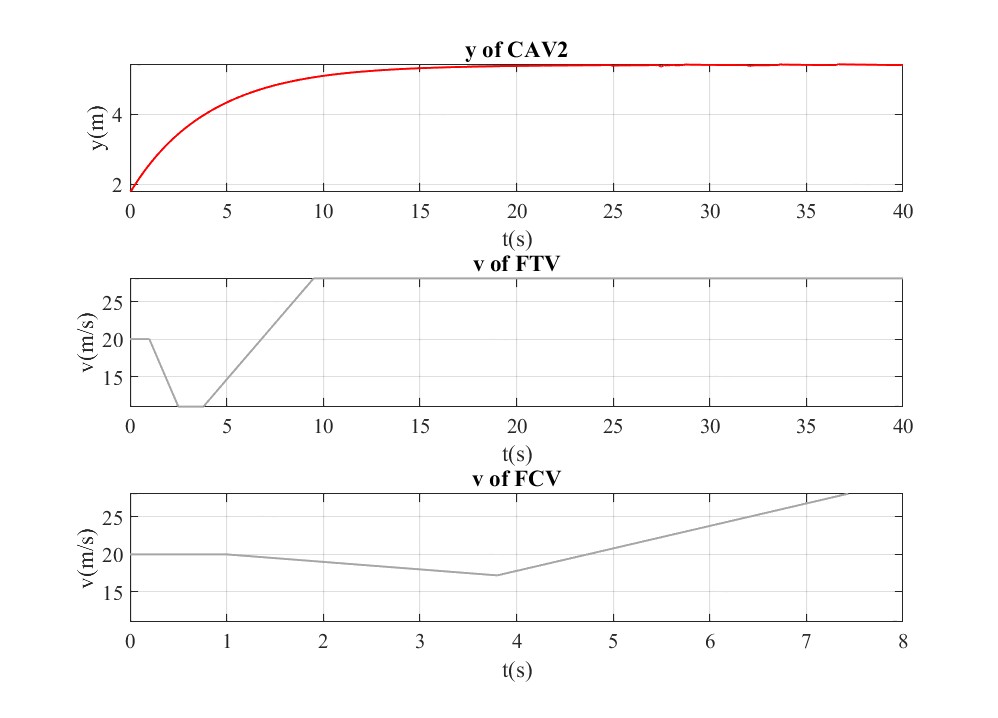}}
 		\subcaptionbox{Velocities and trajectory in cooperation-based experiment \label{fig:fdec2-withcoop-v-traj}}{\includegraphics[width=9cm]{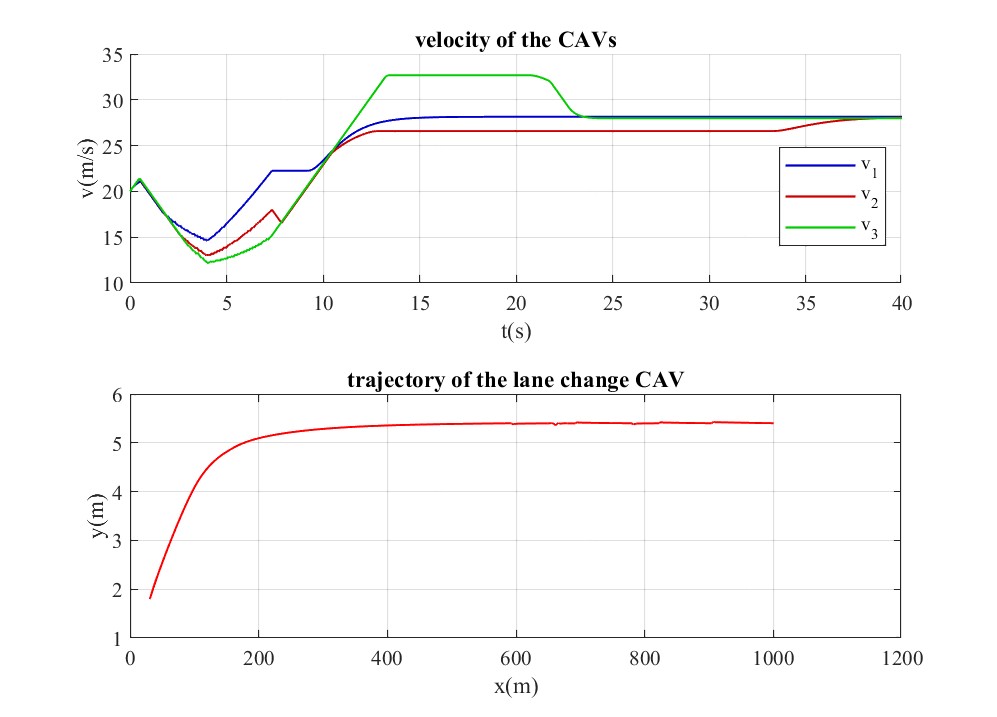}}
 		\quad
 		
 		\subcaptionbox{Selected states of key vehicles in experiment without cooperation \label{fig:fdec2-nocoop-collision}}{\includegraphics[width=9cm]{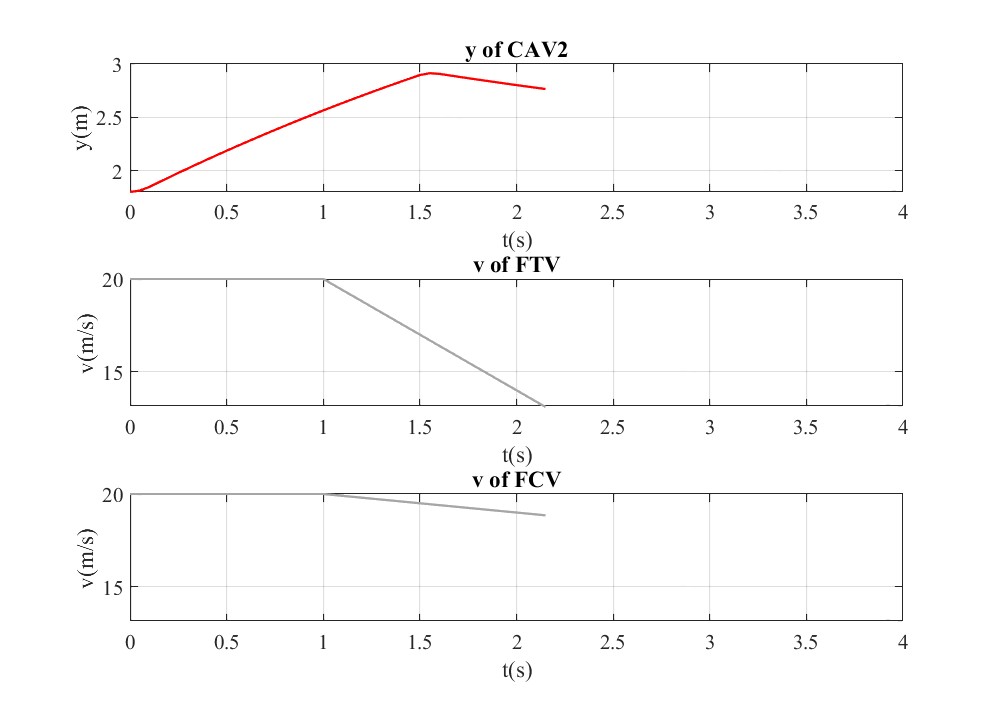}}
 		\subcaptionbox{Velocities and trajectory in experiment without cooperation \label{fig:fdec2-nocoop-v-traj}}{\includegraphics[width=9cm]{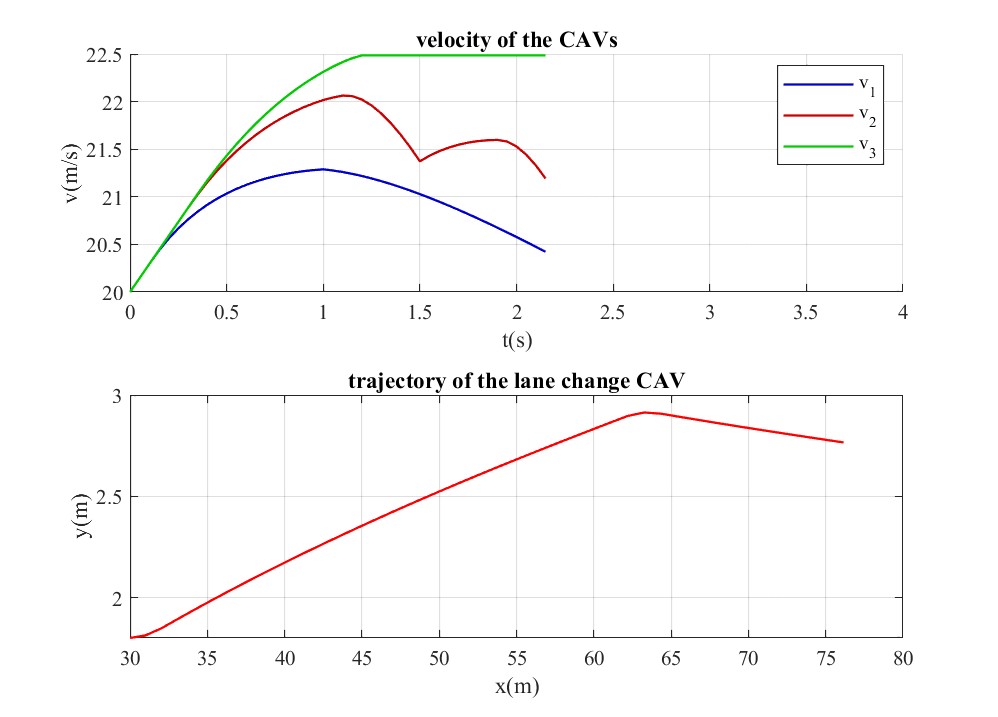}}
 		
 		\caption{Result of comparative experiment in FFDEC scenario}\label{fig:result-fdec2-coop}
 	\end{figure*}

 	The controller parameters are: $\alpha_l=1.7$, $\alpha_y=0.5$, $\alpha_{\psi}=24$, $p_l=15$, $p_y=0.001$, $p_{\psi}=1000$. The result of the CLF-CBF-QP for platoon operation is shown in Fig.\ref{fig:fdec2-withcoop-collision} and Fig.\ref{fig:fdec2-withcoop-v-traj} where the lane change CAV successfully changed its lane. As presented in Fig.\ref{fig:fdec2-withcoop-v-traj} this success was due to the deceleration of the CAV2 and CAV3 at the beginning of the experiment, which was resulted from the split operation of the CAV platoon.
 	
 	The result of the CLF-CBF-QP for single vehicle is shown in Fig.\ref{fig:fdec2-nocoop-collision} and Fig.\ref{fig:fdec2-nocoop-v-traj}. In this experiment, the controller failed to solve the QP problem at $t=2.2\text{s}$ because of the infeasibility. All the three CAVs accelerated at the beginning because they all aimed to reach the desired speed driven by CLF for single vehicle. As the FTV and the FCV decelerated, the lane change CAV (CAV2) tried to turn back to avoid the potential collision between the FTV and itself. Shown in \ref{fig:fdec2-nocoop-v-traj}, at the same time, however, the CAV1 decelerates to avoid the potential collision between the CAV1 and the FCV. The CAV3 did not decelerate because it was following CAV1 at the moment without platoon operation. Then the spacing gap between the CAV1 and the CAV3 actually decreased. Finally, the CBF constraints of the CAV2 could not be satisfied, and the QP problem failed.

	\section{Conclusion}
	In this paper, an optimization-based controller using CLFs and CBFs is developed for platoon operation and safety-critical lane change control. Four safety-critical scenarios are defined for platoon lane change problems. Based on numerical simulation, comparative experiments are conducted to show the effectiveness of the proposed controller. Compared to CLF-QP controller, the proposed CLF-CBF-QP controller can not only accomplish the desired platoon operation, but also guarantee safety when doing lane change. The designed cooperative platoon operation can also facilitate the lane change by adjusting the headway between CAVs. In the future work, communication delay will be considered, and more scenarios will be involved to test the performance of CBF-based controller.

    \bibliography{ref_all}
    
\end{document}